# General 2-path Problem

Qianghui Xiao[a]

[a] *College of Electrical and Information Engineering, Hunan University of Technology, Zhuzhou, Hunan, 412007, PR China*
*E-mail address: 10010@hut.edu.cn;qh.xiao004@163.com*

**Abstract**

In this paper, some preliminaries about signal flow graph, linear time-invariant system on $F(z)$ and computational complexity are first introduced in detail. In order to synthesize the necessary and sufficient condition on $F(z)$ for a general 2-path problem, the sufficient condition on $F(z)$ or $R$ and necessary conditions on $F(z)$ for a general 2-path problem are secondly analyzed respectively. Moreover, an equivalent sufficient and necessary condition on $R$ whether there exists a general 2-path is deduced in detail. Finally, the computational complexity of the algorithm for this equivalent sufficient and necessary condition is introduced so that it means that the general 2-path problem is a P problem.

## 1. Introduction

During the last 50 years, a great deal of interest has been brought to the problem of computational complexity of the algorithms. Since S. A. Cook puts forward the nondeterministic polynomial time (referred as NP) complete problem known as NPC problem (Cook, 1971, 2000; Bang & Gutin, 2002), then R. M. Karp used these completeness results to show that 20 other natural problems are NP-complete, thus forcefully demonstrating the importance of the subject (Karp, 1972; Cook, 2000). Karp also introduced the now standard notation P and NP and redefined NP-completeness using the polynomial time analog of many-one reducibility, a definition that has become standard (Karp, 1972; Cook, 2000). Meanwhile L. A. Levin, independently of Cook and Karp, defined the notion of "universal search problem", similar to the NP-complete problem, and gave six examples, including Satisfiability (Levin, 1973; Cook, 2000).

More than 50 years have passed, and there are many arguments for P versus NP, but it remains an open problem.

2-path problem is a typical NPC decision problem (Fortune, Hopcroft, & Wyllie, 1980; Bang & Gutin, 2002). The general 2-path problem is a generalized treatment for a 2-path problem with the different mathematic permutations for its control input indexes, so of course it is also a NPC decision problem. One possible non-constructive proof method (i.e., existence proof method) for the standard NP-complete problems stated in the official problem statement (Cook, 2000) will be used to analyze the general 2-path problem.

This paper first points out that there is a one-to-one correspondence between the linear time-invariant system and signal flow graph without any parallel branches, so that the signal flow graph combined with its corresponding linear time-invariant system can be used to obtained one necessary and sufficient condition on $R$ or $F(z)$ whether there exists a general 2-path.

The paper is organized as follows: the next section is devoted to some preliminaries in detail. Section 3 explains a sufficient condition on $R$ or $F(z)$ whether there exists a general 2-path. Section 4 analyzes a necessary condition on $F(z)$ whether there is a general 2-path, and then section 5 deduces the sufficient and necessary condition on $F(z)$. Moreover, an equivalent sufficient and necessary condition on $R$ whether there exists a general 2-path is deduced in detail. Finally, section 8 introduces the computational complexity of the algorithm for this equivalent sufficient and necessary condition.

In this paper, the general 2-path problem is considered and analyzed by default, and the analysis of linear time-invariant systems is based on the zero-state response characteristics which will not be mentioned later.

## 2. Preliminaries

**Definition 2.1.** The symbol $x$ represents not only the column vector of state variables (denoted as $x = [x_1, x_2, \cdots, x_n]^T$), but also the set of state variables (denoted as $x = \{x_i\}_n$), and at the same time the set of first-order branches in the signal flow graph corresponding to the linear system.

**Definition 2.2.** The symbol $x_i$ represents not only the state variable, but also the first-order branch with $x_i$ as its output, and at the same time the output of the first-order branch in the signal flow graph corresponding to the system.

**Definition 2.3.** The symbol $y$ represents not only the column vector of outputs (denoted as $y = [y_1, y_2]^T$), but also the set of outputs (denoted as $y = \{y_1, y_2\}$).

**Definition 2.4.** The symbol $u$ represents not only the column vector of control inputs (denoted as $u = [u_1, u_2]^T$), but also the set of control inputs (denoted as $u = \{u_1, u_2\}$).

### 2.1. Fundamentals about signal flow graph

The signal flow graph (often referred as SFG) originated from Mason's use of graphical method to describe linear algebraic equations. It is a signal transmission network composed of nodes and directed branches (referred as branches), and is a typical application of directed graph (or digraph). Based on the zero-state response, a linear system can be converted into a corresponding linear algebraic system by using Laplace transform method, so that a signal flow graph can be used to describe the linear system. Linear system and signal flow graph can be converted to each other, so that there is a one-to-one correspondence between the two (S. J. Mason, 1956; Bang & Gutin, 2002).

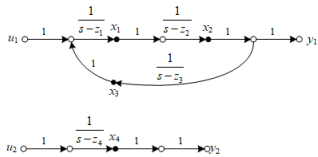

Fig.1. A standard signal flow graph

It is assumed that the branches in signal flow graphs are divided into two types: general first-order branch and zero-order branch with unity gain. One general first-order branch consists of a first-order branch (its input node is the input node of the general first-order branch, and its output is the state variable corresponding to this first-order branch) with a gain of $1/(s-z_i)$ (i.e., the open-loop pole of this first-order branch such as $x_i$ is one indeterminate $z_i$) and a zero-order branch with unity gain (i.e., combination coefficient of this first-order branch) in series. It is no problem to assume that each indeterminate as above is independent of time so that the obtained linear system corresponding to a signal flow graph is time-invariant. A standard signal flow graph is shown in Fig.1.

In the following part, it is assumed that one directed graph (or digraph) has been transformed into a standard signal flow graph without any parallel branches that the open-loop pole of each first-order branch (such as $x_i$) is one indeterminate $z_i$ and the combination coefficient of this first-order branch is one.

**Definition 2.5.** The symbol $z$ represents not only the column vector of all open-loop poles (denoted as $z = [z_1, z_2, \cdots, z_n]^T$), but also the set of the open-loop poles (denoted as $z = \{z_i\}_n$).

**Definition 2.6.** The symbol $z_i$ represents not only the open-loop pole of the first-order branch $x_i$, but also the first-order branch $x_i$.

**Definition 2.7.** A route from a node (i.e., input node such as $v_1$) to another node (i.e., output node such as $v_2$) is called a walk (denoted as $walk(v_1, v_2)$), and a walk that passes through its each node only once is called a path (denoted as $path(v_1, v_2)$). For a linear system, one path from a control input node (such as $u_i$) to an output node (such as $y_j$) is called a forward path (denoted as $path(u_i, y_j)$), and all forward paths from $u_i$ to $y_j$ are denoted as $(u_i, y_j)$. Multiple paths without common nodes between each other are called node-disjoint paths or non-contact paths.

**Definition 2.8.** A closed path with its start and end points at the same node is called a loop or cycle. Multiple loops without common nodes between each other are called node-disjoint loops or non-contact loops. A loop that intersects a first-order branch with its input node is called a contact loop or node-joint loop of this first-order branch. Then a loop that intersects a first-order branch with its output node is called a branch-joint loop of this first-order branch (i.e., the first-order branch is included in this loop).

A loop set is called an associative loop set when each first-order branch included in this loop set is connected to another first-order branch by one path in the loop set. An associative loop set is called the associative loop set of a certain first-order branch when the first-order branch and this associative loop set are branch-joint (i.e., the first-order branch is included in this associative loop set). If no node intersects between two loop sets (especially two associative loop sets), they are called two node-disjoint loop sets (especially two node-disjoint associative loop sets).

**Definition 2.9.** The gain of a walk such as $w_i$ (or path $p_i$, or loop $L_i$) is defined as the product of the gains of all first-order branches contained in this walk (or path, or loop), denoted as $w_i$ or $p_i$, or $L_i$.

**Definition 2.10.** The symbol $w_i$ (or $p_i$, or $L_i$) represents not only the gain of a walk $w_i$ (or path $p_i$, or loop $L_i$), but also the set of all first-order branches in this walk $w_i$ (or path $p_i$, or loop $L_i$).

**Definition 2.11.** The order or length of a walk $w_i$ (or path $p_i$, or loop $L_i$) is defined as the number of first-order branches contained in this walk (or path, or loop), denoted as $len(w_i)$ (or $len(p_i)$, or $len(L_i)$).

**Proposition 2.12.** For a linear system, if there is at least one forward path from a control input node (such as $u_i$) to an output node (such as $y_j$), then there exists at least one path with the shortest length (i.e., shortest path denoted as $path(u_i, y_j)_{\min}$, whose length is denoted as $len(u_i, y_j)_{\min}$), and all shortest paths from $u_i$ to $y_j$ are denoted as $(u_i, y_j)_{\min}$.

**Proposition 2.13.** The closed-loop pole of each first-order branch is determined by its associative loop set.

**Definition 2.14.** A branch whose input node and output node are the same node is called self-loop. Without any loss of generality, it is assumed that there is no self-loop included in the signal flow graphs studied in this paper.

**Definition 2.15.** The in-degree of the node $v$ in a signal flow graph is defined as the total number of branches whose output nodes are the node $v$ (denoted as $d^-(v)$); the out-degree of the node $v$ is defined as the total number of branches whose input nodes are the node $v$ (denoted as $d^+(v)$). For a input node $u_i$, $d^-(u_i) = 0$; and for a output node $y_j$, $d^+(y_j) = 0$.

**Definition 2.16.** If the in-degree and out-degree of a node $v$ are all more than one, a splitting operation for the node $v$ can be used to replace the node $v$ by a branch such as $(v_1, v_2)$ composed of two new nodes (i.e., $v_1$ and $v_2$), which an original branch such as $(w_1, v)$ becomes $(w_1, v_1)$ and another original branch such as $(v, w_2)$ becomes $(v_2, w_2)$.

After the splitting operations for the nodes whose in-degree and out-degree are all more than one, at least one between the in-degree and out-degree of any node except the input nodes and output nodes is not more than one (i.e., that is one or zero). Therefore, the node-joint paths (or loops) are also branch-joint, and vice versa.

Without any loss of generality, it is assumed that at least one between the in-degree and out-degree of any node except the input nodes and output nodes is not more than one in the signal flow graphs studied in this paper.

**Definition 2.17.** The order of a signal flow graph $D$ is defined as the total number of all nodes in the signal flow graph (denoted as $|D|$). And the size of a signal flow graph $D$ is defined as the total number of all first-order branches in the signal flow graph (denoted as $\|D\|$). The maximum absolute value of the open-loop poles in a signal flow graph is denoted as $|w_{\max}|$. Then, $|D|$, $\|D\|$ and $\log_2 |w_{\max}|$ will be used to measure the computational complexity of the subsequent decision algorithm whether there exists a general 2-path (Bang & Gutin, 2002).

**Definition 2.18.** 2-path problem can be stated as: for a given signal flow graph and its different nodes such as $u_1, u_2, y_1, y_2$, whether there is two node-disjoint paths such as $P_1, P_2$ in this signal flow graph, so that each path $P_i$ is $path(u_i, y_i), i = 1, 2$ (Bang & Gutin, 2002)?

**Definition 2.19.** General 2-path problem can be stated as: for a given signal flow graph and its different nodes such as $u_1, u_2, y_1, y_2$, whether there is two node-disjoint paths such as $P_1, P_2$ (called a general 2-path) in this signal flow graph, so that each path $P_i$ (called a general path $P_i$) is $path(u_{j_i}, y_i)$, $i = 1, 2$? The sequence $j_1, j_2$ is a kind of mathematic permutation of $1, 2$.

The general 2-path problem is a generalized treatment for a 2-path problem with the different mathematic permutations for its control input indexes.

**Proposition 2.20.** The Mason's gain formula is as follows (S. J. Mason, 1956).

$$G(s) = \frac{v_{out}}{v_{in}} = \sum_{\mu=1}^{N} p_\mu \frac{\Delta_\mu}{\Delta} \quad (1)$$

$$\Delta = 1 - \sum L_i + \sum L_i L_j - \sum L_i L_j L_k + \cdots \quad (2)$$

where

$\Delta$: the determinant of the given signal flow graph.
$v_{in}$: input-node variable.
$v_{out}$: output-node variable.

$G(s)$: complete gain (or total gain, or transfer function) between $v_{in}$ and $v_{out}$.
$N$: total number of forward paths between $v_{in}$ and $v_{out}$.
$p_\mu$: path gain of the $\mu th$ forward path between $v_{in}$ and $v_{out}$.
$L_i$: loop gain of each closed loop in the graph.
$L_i L_j$: product of the loop gains of any two node-disjoint loops in the graph.
$L_i L_j L_k$: product of the loop gains of any three pairwise node-disjoint loops in the graph.
$\Delta_\mu$: the cofactor value of $\Delta$ for the $\mu th$ forward path, with the loops node-jointing the $\mu th$ forward path removed.

For a given signal flow graph whose first-order branches' open-loop poles are indeterminates, the irreducible polynomials (each of which corresponds to a unique associative loop set) in the characteristic polynomial of its corresponding linear system will be obtained in the following, which can be used to analyze the subsequent necessary condition on $F(z)$ whether there is a general 2-path.

**Theorem 2.21.** For a given signal flow graph whose first-order branches' open-loop poles are indeterminates, it is assumed that there are only $M$ node-disjoint associative loop sets (denoted as $Q_1, Q_2, \cdots, Q_M$) in the graph and their corresponding sub-determinants (i.e., the determinants of these associative loop sets whose expression forms are the same as that of $\Delta$) are $\overline{\Delta}_1, \overline{\Delta}_2, \cdots, \overline{\Delta}_M$ respectively. Then, the sub-determinants are all irreducible factors about loop gains (or first-order branch gains) in the determinant of the graph, and the following identity holds.

$$\Delta = \prod_{i=1}^{M} \overline{\Delta}_i \quad (3)$$

**Proof.** It is obvious that the equation (3) holds.

1) It is assumed that the sub-determinant $\overline{\Delta}_i$ about loop gains is reducible and $\overline{\Delta}_i = \overline{\Delta}_{i_1} \cdot \overline{\Delta}_{i_2}$, so the expression form of $\overline{\Delta}_{i_1}$ (or $\overline{\Delta}_{i_2}$) can only be the same as that of $\overline{\Delta}_i$ (or $\Delta$). The loop set corresponding to $\overline{\Delta}_{i_1}$ (or $\overline{\Delta}_{i_2}$) is assumed to be $Q_{i_1}$ (or $Q_{i_2}$). ⇒ Any one loop (in $Q_i$) is either in $Q_{i_1}$ or in $Q_{i_2}$. ⇒ For any loop such as $L_\mu$ (or $L_\nu$) in $Q_{i_1}$ (or $Q_{i_2}$), there is the product item of $L_\mu$ and $L_\nu$ in $\overline{\Delta}_i$, so that $Q_{i_1}$ and $Q_{i_2}$ are node-disjoint according to equation (2). ⇒ $Q_{i_1} \cap Q_{i_2} = \varnothing$ ⇒ It means that for any branch such as $x_j$ (or $x_k$) in $Q_{i_1}$ (or $Q_{i_2}$), there is no path between $x_j$ and $x_k$ in the associative loop set $Q_i$, so that it is contradictory to the definition of an associative loop set (see Definition 2.8). Therefore, the sub-determinants are all irreducible factors about loop gains in the determinant of the graph.

2) It is assumed that the sub-determinant $\overline{\Delta}_i$ about first-order branch gains is reducible and $\overline{\Delta}_i = \overline{\Delta}_{i_1} \cdot \overline{\Delta}_{i_2}$. ⇒ It can be deduced from the expression form of $\overline{\Delta}_i$ that each first-order branch gain (so that this first-order branch) is contained either in $\overline{\Delta}_{i_1}$ or in $\overline{\Delta}_{i_2}$. ⇒ For any loop such as $L_j$ in the associative loop set $Q_i$, any one first-order branch such as $x_k$ (it is no problem to assume that this branch gain is only in $\overline{\Delta}_{i_1}$) in the loop $L_j$ is selected for the opening operation (i.e., the open pole $z_k \to \infty$ ⇒ the branch gain $(s - z_k)^{-1} \to 0$). ⇒ Then the items that contain $(s - z_k)^{-1}$ in $\overline{\Delta}_{i_1}$ become zero but $\overline{\Delta}_{i_2}$ does not change, and at the same time, all loop gains which contain $(s - z_k)^{-1}$ also become zero. ⇒ It means that all loop gains which con-tain $(s-z_k)^{-1}$ are in $\bar{\Delta}_{i_1}$ but not in $\bar{\Delta}_{i_2}$. $\Rightarrow$ So the gains of all first-order branches in anyone loop are either in $\bar{\Delta}_{i_1}$ or in $\bar{\Delta}_{i_2}$. $\Rightarrow$ According to the expression form of $\bar{\Delta}_i$, the expression form of $\bar{\Delta}_{i_1}$ (or $\bar{\Delta}_{i_2}$) can only be the same as that of $\bar{\Delta}_i$ (or $\Delta$), in which each loop gain is replaced by the product of the gains of all first-order branches in the loop. The loop set corresponding to $\bar{\Delta}_{i_1}$ (or $\bar{\Delta}_{i_2}$) is assumed to be $Q_{i_1}$ (or $Q_{i_2}$). $\Rightarrow$ $Q_{i_1} \cap Q_{i_2} = \varnothing$ $\Rightarrow$ As the same derivation as above, there exists a contradiction to the definition of an associative loop set (see Definition 2.8). Therefore, the sub-determinants are all irreducible factors about first-order branch gains in the determinant of the graph. $\square$

**Definition 2.22.** The primary factor of a first-order branch is defined as the inverse of its gain, referred as the primary factor.

The following theorem will be easy deduced based on Theorem 2.21.

**Theorem 2.23.** For a given signal flow graph (whose first-order branches' open-loop poles are indeterminates) and its corresponding linear system, it is assumed that there are only $M$ node-disjoint associative loop sets in the graph and the characteristic polynomial of the linear system is $\Delta(s)$. Then $\Delta(s)$ can be decomposed into $M$ irreducible polynomial factors (each of which corresponds to one unique associative loop set) about primary factors of the first-order branches (referred as irreducible polynomials) and other primary factors of the first-order branches that are not in any loop.

### 2.2. Linear time-invariant system on $F(z)$

**Definition 2.24.** Let $z_1, z_2, \cdots, z_n$ denote $n$ indeterminates (can also be independent parameters), not constants or numerical values. Let $z = [z_1, z_2, \cdots, z_n]^T \in R^n$, $R^n$ is the domain of definition for $z$, and it can also be called parameter space. Let $F(z)$ denote the field of all rational functions with real coefficients in $n$ indeterminates $z_1, z_2, \cdots, z_n$, and $F(z)[s]$ denote the ring of all $F(z)$-coefficient polynomials in $s$ (K. S. Lu, 2012).

**Definition 2.25.** If any entry of matrix is a member of $F(z)$ (i.e., rational function in $z_1, z_2, \cdots, z_n$), then this matrix is called a rational function matrix (referred as RFM) in $z$ or a matrix on $F(z)$; if the coefficient matrices of a linear system are considered to be RFMs, the system is called the rational function system in multi-parameters, simply called rational function system (referred as RFS), or system on $F(z)$ (K. S. Lu, 2012).

**Proposition 2.26.** If $f(z) \in F(z)$, then either $f(z) \equiv 0$ or $m^*\{z \in R^n | f(z) = 0\} = 0$ (K. S. Lu, 2012), that is to say, $f(z) \not\equiv 0 \Leftrightarrow m^*\{z \in R^n | f(z) = 0\} = 0$.

where $m^*\{\cdot\}$ denotes the Lebesgue measure of the set $\{\cdot\}$.

If $f(z) \equiv 0$ on $F(z)$,
$$\Phi_0 \triangleq \{z \in R^n | f(z) \equiv 0\} = R^n \tag{4}$$

If $f(z) \not\equiv 0$ on $F(z)$,
$$\Phi_1 \triangleq \{z \in R^n | m^*\{f(z) = 0\} = 0\} \tag{5}$$
$$\Phi_2 \triangleq \{z \in R^n | f(z) \neq 0\} \tag{6}$$

$$\begin{cases} \Phi_1 \cap \Phi_2 = \varnothing \\ \Phi_1 \cup \Phi_2 = R^n \end{cases} \tag{7}$$

The set $\Phi_1$ is a singular set, which is the essential cause of the difference between the necessary condition on $F(z)$ and the necessary condition on $R$ for a general 2-path problem.

**Theorem 2.27.** For $f(z) \in F(z)$, if there exists one value $\alpha \in R^n$ such that $f(\alpha) \neq 0$, then $f(z) \not\equiv 0$.

**Proof.** It is assumed that $f(z) \equiv 0$, then $f(\alpha) = 0$. It is a contradiction. So Theorem 2.27 holds. $\square$

**Definition 2.28.** For a given standard signal flow graph (whose first-order branches' open-loop poles are indeterminates), its corresponding linear time-invariant system is system $\Sigma(A, B, C)$ with two inputs and two outputs (referred as TITO) given by

$$\begin{cases} \dot{x} = Ax + Bu \\ y = Cx \end{cases} \tag{8}$$

$$A = \begin{bmatrix} z_1 & * & \cdots & * & * \\ * & z_2 & \ddots & * & * \\ \vdots & \vdots & \ddots & \ddots & \vdots \\ * & * & \cdots & z_{n-1} & * \\ * & * & \cdots & * & z_n \end{bmatrix} \triangleq A_0 + A_z \tag{9}$$

$$A_z = diag[z_1, z_2, \cdots, z_n] \tag{10}$$

$$\begin{cases} B = [b_1, b_2] \\ C = \begin{bmatrix} c_1 \\ c_2 \end{bmatrix} \end{cases} \tag{11}$$

$$\frac{\partial A}{\partial z_i} = \frac{\partial A_z}{\partial z_i} = e_i e_i^T, i = 1, 2, \cdots, n \tag{12}$$

$$I = diag[1, 1, \cdots, 1] \triangleq [e_1, e_2, \cdots, e_n] \tag{13}$$

where the symbol "*" in the matrix $A$ represents the entry that is either one or zero, and any element in $B$ (or $C$) is also either one or zero. $x = [x_1, x_2, \cdots, x_n]^T \in R^n$, $u = [u_1, u_2]^T \in R^2$, $y = [y_1, y_2]^T \in R^2$.

**Proposition 2.29.** For a TITO linear time-invariant system $\Sigma(A, B, C)$ as shown in equations (8), then its transfer function is shown in the following (W. A. Wolovich, 1974; F. S. Pang, 1992).

$$T(s, z) = C(sI - A)^{-1} B \tag{14}$$

$$T(s, z) = \sum_{i=0}^{\infty} \frac{CA^i B}{s^{i+1}} \tag{15}$$

$$T(s, z) \equiv 0 \Leftrightarrow rank T(s, z) = 0 \tag{16}$$

**Definition 2.30.** Let $d_i$ be given by
$$d_i = \min\{j | c_i A^{j-1} B \neq 0, j = 1, 2, \cdots, n\} \tag{17}$$
or
$$d_i = n \quad \text{if } c_i A^{j-1} B = 0 \text{ for all } j \tag{18}$$

where $c_i$ denotes the *ith* row of $C$, and integer $d_i$ is defined as the relative order corresponding to the output $y_i$.

The following SISO subsystem with only one path from $u_1$ to $y_1$ (i.e., shortest path, whose order or length is $d$ (that is relative order)) as an example is considered to analyze the relationship between the first three coefficients of the series expansion on the right side of equation (15) and the indeterminates, so that the results will be generalized to the SISO subsystem with multiple paths. It is assumed that $z_1, z_2, \cdots, z_d$ denote the open poles of $d$ first-order branches in sequence in $path(u_1, y_1)$ (or $(u_1, y_1)$).

$$c_1(sI-A)^{-1}b_1 = \frac{c_1 A^{d-1} b_1}{s^d} + \frac{c_1 A^d b_1}{s^{d+1}} + \frac{c_1 A^{d+1} b_1}{s^{d+2}} + \cdots \quad (19)$$

For the standard signal flow graph and its corresponding linear system (as shown in(8)), the following equations hold.

$$\begin{cases} c_1 A^{j-1} b_1 = 0, 1 \le j < d \\ c_1 A^{d-1} b_1 = 1 \triangleq f_0(z) \end{cases} \quad (20)$$

For anyone first-order branch such as $x_i$ (whose input node is denoted as $u(x_i)$) in the path $(u_1, y_1)$, it is no problem to assume that the order of $(u_1, x_i)$ is $\bar{k}_i$ and the order of $(u(x_i), y_1)$ is $k_i$. The path $(u_1, y_1)$ can be decomposed as shown in Fig.2.

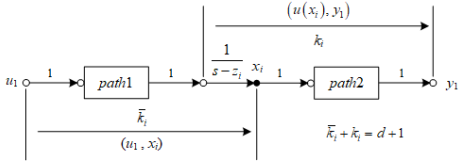

Fig.2 The schematic diagram of path $(u_1, y_1)$

$$\begin{cases} e_i^T A^{j-1} b_1 = 0, 1 \le j < \bar{k}_i \\ e_i^T A^{\bar{k}_i - 1} b_1 = 1 \end{cases} \quad (21)$$

$$\begin{cases} c_1 A^{j-1} e_i = 0, 1 \le j < k_i \\ c_1 A^{k_i - 1} e_i = 1 \end{cases} \quad (22)$$

So

$$\frac{\partial}{\partial z_i}(c_1 A^d b_1) = \sum_{v=1}^d (c_1 A^{v-1})(\frac{\partial A}{\partial z_i})(A^{d-v} b_1), i = 1, 2, \cdots, d \quad (23)$$

Substitute (12) into (23), utilize (21) and (22), and then obtain

$$\frac{\partial}{\partial z_i}(c_1 A^d b_1) = (c_1 A^{k_i - 1} e_i)(e_i^T A^{\bar{k}_i - 1} b_1) = 1, i = 1, 2, \cdots, d \quad (24)$$

$$c_1 A^d b_1 = \sum_{v=1}^d z_v + f_{10}(z) \quad (25)$$

where $f_{10}(z)$ is a constant.

$$y_1^{(d+1)} = c_1 A^{d+1} x + c_1 A^{d-1} b_1 u_1^{(1)} + c_1 A^d b_1 u_1 \quad (26)$$

For the SISO subsystem with only one path $(u_1, y_1)$, if $(z_1, z_2, \cdots, z_d) = 0$, the following equation holds.

$$c_1 A^d b_1 u_1 = 0, (z_1, z_2, \cdots, z_d) = 0 \quad (27)$$

Therefore

$$c_1 A^d b_1 = 0, (z_1, z_2, \cdots, z_d) = 0 \quad (28)$$

$$f_{10}(z) = 0 \quad (29)$$

$$c_1 A^d b_1 = \sum_{v=1}^d z_v \triangleq f_{11}(z) \quad (30)$$

Based on the same derivations as above, the following equations hold.

$$\begin{cases} e_i^T A^{\bar{k}_i} b_1 = \sum_{v=1}^i z_v \\ c_1 A^{k_i} e_i = \sum_{v=i}^d z_v \end{cases} \quad (31)$$

$$\frac{\partial}{\partial z_i}(c_1 A^{d+1} b_1) = \sum_{v=1}^{d+1}(c_1 A^{v-1})(\frac{\partial A}{\partial z_i})(A^{d+1-v} b_1), i = 1, 2, \cdots, d \quad (32)$$

$$\frac{\partial}{\partial z_i}(c_1 A^{d+1} b_1) = (c_1 A^{k_i - 1} e_i)(e_i^T A^{\bar{k}_i} b_1) + (c_1 A^{k_i} e_i)(e_i^T A^{\bar{k}_i - 1} b_1) \quad (33)$$

$$\frac{\partial}{\partial z_i}(c_1 A^{d+1} b_1) = e_i^T A^{\bar{k}_i} b_1 + c_1 A^{k_i} e_i = z_i + \sum_{v=1}^d z_v \quad (34)$$

$$\frac{\partial^2}{\partial z_i^2}(c_1 A^{d+1} b_1) = \frac{\partial}{\partial z_i}(z_i + \sum_{v=1}^d z_v) = 2 \quad (35)$$

$$c_1 A^{d+1} b_1 = \sum_{v=1}^d z_v^2 + \sum_{1 \le i < j \le d} z_i z_j + f_{21}(z) + f_{20}(z) \quad (36)$$

where $f_{20}(z)$ is a constant and $f_{21}(z)$ is a primary polynomial about the $d$ indeterminates.

$$\frac{\partial}{\partial z_i} f_{21}(z) = 0, i = 1, 2, \cdots, d \quad (37)$$

$$f_{21}(z) = 0 \quad (38)$$

$$y_1^{(d+2)} = c_1 A^{d+2} x + c_1 A^{d-1} b_1 u_1^{(2)} + c_1 A^d b_1 u_1^{(1)} + c_1 A^{d+1} b_1 u_1 \quad (39)$$

For the SISO subsystem with only one path $(u_1, y_1)$, if $(z_1, z_2, \cdots, z_d) = 0$, the following equations hold.

$$\begin{cases} c_1 A^{d+1} b_1 u_1 = 0, (z_1, z_2, \cdots, z_d) = 0 \\ c_1 A^d b_1 u_1^{(1)} = 0, (z_1, z_2, \cdots, z_d) = 0 \end{cases} \quad (40)$$

So

$$c_1 A^{d+1} b_1 = 0, (z_1, z_2, \cdots, z_d) = 0 \quad (41)$$

$$f_{20}(z) = 0 \quad (42)$$

$$c_1 A^{d+1} b_1 = \sum_{v=1}^d z_v^2 + \sum_{1 \le i < j \le d} z_i z_j \triangleq f_{22}(z) \quad (43)$$

**Definition 2.31.** For a SISO subsystem with only one $path(u_1, y_1)$, $z_1, z_2, \cdots, z_d$ denote the open poles of $d$ first-order branches in sequence in $path(u_1, y_1)$, then $f_{22}(z)$ (as shown in (43)) is called elementary symmetric quadratic polynomial of the path about the $d$ indeterminates (referred as ESQP of the path), $f_{11}(z)$ (as shown in (30)) is called elementary symmetric primary polynomial of the path (referred as ESPP of the path), and $f_0(z)$ (as shown in (20)) is called the total number of the path.

**Theorem 2.32.** For a standard signal flow graph, if there is only one $path(u_1, y_1)_{\min}$ whose length is $d$ in $(u_1, y_1)$, $f_{22}(z)$ (as shown in (43)) is the ESQP of the shortest path.

**Theorem 2.33.** For a standard signal flow graph, if there is $f_{22}(z)$ (as shown in (43), which is the ESQP about the $d$ indeterminates) in the third item of the series expansion for $c_1(sI-A)^{-1}b_1$, there exists one $path(u_1, y_1)_{\min}$ with the length of $d$ in $(u_1, y_1)$, which includes the $d$ first-order branches corresponding to these $d$ indeterminates.

**Proof.** Any item of $z_i z_j (\forall i \ne j)$ in $f_{22}(z)$ means that two branches $x_i$ and $x_j$ must be included in the same shortest path in $(u_1, y_1)$. It is obvious that the lengths of the shortest paths, each of which is respectively from the input node $u_1$ to the output node of each first-order branch whose corresponding indeterminate is all contained in $f_{22}(z)$, are not equal to each other and must be a kind of mathematical permutation of $1, 2, \cdots, d$.

It is no problem to assume that the length of the shortest path from the input node $u_1$ to the output node of the branch $x_{i_j}$ is $j$ (i.e., $len(u_1, x_{i_j})_{\min} = j, j = 1, 2, \cdots, d$).

Any item of $z_{i_j} z_{i_{j+1}} (j = 1, 2, \cdots, d-1)$ in $f_{22}(z)$ means that two branches $x_{i_j}$ and $x_{j+1}$ must be included in the same shortest path in $(u_1, y_1)$ and the input node of branch $x_{i_{j+1}}$ is connected to the output node of branch $x_{i_j}$ through a zero-order branch with unity gain. Furthermore, the input node of branch $x_{i_1}$ is connected to the input node $u_1$ through a zero-order branch with unity gain and the output node of branch $x_{i_d}$ is connected to the output node $y_1$ through a zero-order branch with unity gain. Therefore, there must be one shortest path in $(u_1, y_1)$, which is composed of the first-order branches $x_{i_1}, x_{i_2}, \cdots, x_{i_d}$ in turn. That is to say, this shortest path with the length of $d$ in $(u_1, y_1)_{\min}$ must be composed of these $d$ branches whose corresponding open poles are included in $f_{22}(z)$. □

**Theorem 2.34.** For a standard signal flow graph and its system $\Sigma(A, B, C)$, there is one $path(u_1, y_1)_{\min}$ with the length of $d$ in $(u_1, y_1)$ if and only if there is the ESQP of this shortest path in the third item of the series expansion for $c_1(sI - A)^{-1}b_1$.

For a SISO subsystem with multiple paths, the following theorem can be obtained by use of the linear superposition principle.

**Definition 2.35.** For a standard signal flow graph (whose corresponding linear system is shown in (8)) and its SISO subsystem with multiple paths between the input $u_1$ and output $y_1$, if the order of each shortest path (i.e., relative order) is $d$, let

$$\begin{cases} c_1 A^{j-1} b_1 = 0, 1 \le j < d \\ c_1 A^{d-1} b_1 = f_0(z) \end{cases} \quad (44)$$

$$c_1 A^d b_1 = f_{11}(z) + f_{10}(z) \quad (45)$$

$$c_1 A^{d+1} b_1 = f_{22}(z) + f_{21}(z) + f_{20}(z) \quad (46)$$

Then $f_0(z)$ is the total number of all shortest paths, $f_{11}(z)$ is the sum of the ESPPs of all shortest paths, and $f_{22}(z)$ is the sum of the ESQPs of all shortest paths; $f_{10}(z)$ is the total number of all paths (or walks) whose orders are $(d+1)$ and $f_{21}(z)$ is the sum of the ESPPs of all paths (or walks) whose orders are $(d+1)$; $f_{20}(z)$ is the total number of all paths (or walks) whose orders are $(d+2)$.

**Theorem 2.36.** For a standard signal flow graph and its system $\Sigma(A, B, C)$, there are $n_{11}$ shortest paths whose order is $d$ in $(u_1, y_1)$ if and only if the function $f_{22}(z)$ (as shown in (46)) can be uniquely decomposed into $n_{11}$ ESQPs (as shown in (43)), each of which uniquely corresponds to one unique shortest path in $(u_1, y_1)$. Therefore, the function $f_{22}(z)$ uniquely points to each shortest path in $(u_1, y_1)$ such that also uniquely points to the input node $u_1$ and output node $y_1$.

In the above functions about the open poles of the first-order branches, the function $f_{22}(z)$ is the most important because it contains the bilinear terms of the indeterminates for all shortest paths, which can achieve a one-to-one correspondence between each shortest path in a SISO subsystem and its corresponding ESQP, so that it provides the possibility for transforming the necessary condition on $F(z)$ whether there exists a general 2-path into the necessary condition on $R$.

A square Vandermonde matrix which will be used in the below analysis of the necessary condition on $R$ is shown as follows.

**Proposition 2.37.** A square Vandermonde matrix can be given by

$$V = \begin{bmatrix} 1 & 1 & 1 & \cdots & 1 \\ x_1 & x_2 & x_3 & \cdots & x_m \\ x_1^2 & x_2^2 & x_3^2 & \cdots & x_m^2 \\ \vdots & \vdots & \vdots & \ddots & \vdots \\ x_1^{m-1} & x_2^{m-1} & x_3^{m-1} & \cdots & x_m^{m-1} \end{bmatrix} \quad (47)$$

or

$$V_{i,j} = x_j^{i-1}, 1 \le i \le m, 1 \le j \le m \quad (48)$$

The determinant of a square Vandermonde matrix can be expressed as

$$\det(V) = \prod_{1 \le i < j \le m} (x_j - x_i) \quad (49)$$

This is called the Vandermonde determinant (or Vandermonde polynomial). It is non-zero if and only if all $x_i$ are distinct.

## 2.3. Computational complexity

In the algorithm analysis, time complexity (often referred as complexity) is the very important thing, which reflects the running time of the corresponding computer program on different computers. Because of focusing only on approximations of execution time, the time complexity of an algorithm can be viewed as the number of elementary operations (i.e., arithmetic operations, comparison judgments, data movements, and control triages) of the algorithm. In the case of large-scale operation, the time complexity of an algorithm depends on the input size of the computational object (Bang & Gutin, 2002).

**Definition 2.38.** An algorithm (whose input size is $n$) is called $O(g(n))$ algorithm if the running time of the algorithm with respect to a function $g(n)$ about the input size never exceeds $cg(n)$, where $c$ is constant (that depends only on the algorithm). An algorithm with the complexity $O(g(n))$ is polynomial time algorithm (referred as P) if the function $g(n)$ is a polynomial about $n$ (J. Edmonds, 1965).

For a signal flow graph $D$, its order $|D|$, size $\|D\|$ and $\log_2 |w_{\max}|$ are used to measure its input size so that an algorithm with the complexity $O(g(|D|, \|D\|, \log_2 |w_{\max}|))$ is P if the function $g(|D|, \|D\|, \log_2 |w_{\max}|)$ is a polynomial about $|D|$, $\|D\|$ and $\log_2 |w_{\max}|$ (Bang & Gutin, 2002).

**Definition 2.39.** A problem is called a decision problem, if it asks for a "yes" or "no" answer (Bang & Gutin, 2002).

**Definition 2.40.** A decision problem $S$ is a P problem if and only if there exists a polynomial time algorithm $F$ that finds an answer in the set {"yes", "no"} for any example of $S$ given, such that the answer given by $F$ with input $x$ is "yes" if and only if $x$ is a "yes" example of $S$. The class P is composed of all P problems (Bang & Gutin, 2002).

**Definition 2.41.** A decision problem $S$ is a NP problem if for each "yes" (or "no") example of $S$, there exists a short "proof" (called a certificate) with polynomial size, such that we can use the certificate to verify that it is "yes" (or "no") in polynomial time. The class NP consists of all NP problems (Bang & Gutin, 2002).

**Definition 2.42.** A decision problem $S$ is a NP-hard problem if all problems in NP can be reduced to this problem by use of polynomial-time computable function. If a NP-hard problem also belongs to NP, it is called a NP-complete problem. The class NPC consists of all NP-complete problems (Bang & Gutin, 2002).

**Proposition 2.43.** If a decision problem $S$ is NP-complete and $S \in P$, then $P = NP$ (Cook, 2000).

**Proposition 2.44.** For a TITO signal flow graph (whose first-order branches' open poles are either indeterminates or constant) and its system $\sum(A,B,C)$, 2-path problem on $F(z)$ or $R$ is a NPC decision problem (Bang & Gutin, 2002).

**Proposition 2.45.** For a TITO signal flow graph (whose first-order branches' open poles are either indeterminates or constant) and its system $\sum(A,B,C)$, general 2-path problem on $F(z)$ or $R$ is a NPC decision problem.

## 3. A sufficient condition on $F(z)$ or $R$

It will first be proved below that reversibility of the transfer function matrix $T(s,z)$ or $T(s)$ for a TITO linear system means that there exists at least one general 2-path in the corresponding signal flow graph.

**Theorem 3.1.** For a TITO signal flow graph (whose first-order branches' open poles are either indeterminates or constant) and its system $\sum(A,B,C)$, if its transfer function $T(s,z)$ or $T(s)$ (uniformly denoted as $T(s)$ in the following proof) is invertible, then there exists at least one general 2-path in the signal flow graph.

**Proof.** It is first assumed that there are mutually node-joint forward paths from two control inputs (such as $u_1$ and $u_2$) to two different outputs (such as $y_1$ and $y_2$) though a node (assumed to be $v_0$ which may be a state variable or control input). A row vector composed of the transfer functions of all the forward paths from the two control inputs to the output $y_1$ through the node $v_0$ is $\Delta T_1$, and another row vector composed of the transfer functions of all the forward paths from two control inputs to the output $y_2$ through the same node $v_0$ is $\Delta T_2$.

$$T(s) \triangleq \left[ T_1^T, T_2^T \right]^T \tag{50}$$

It is obvious that $\Delta T_1$ and $\Delta T_2$ are not equal to zero and have a linear relationship with each other. Row transformations are performed on $T(s)$ to eliminate either $\Delta T_1$ or $\Delta T_2$ so that there are no transfer functions of mutually node-joint forward paths from two control inputs to two outputs $y_1$ and $y_2$ though $v_0$ and the invertibility of $T(s)$ is maintained.

$$\Delta T_2 = \lambda \Delta T_1, \lambda \neq 0, \lambda \in R(s) \tag{51}$$

$$\Delta T_1 = \gamma_1 T_1 + \gamma_2 T_2 \tag{52}$$

$$\begin{cases} \hat{T}_1 \triangleq T_1 - \Delta T_1 \\ \hat{T}_2 \triangleq T_2 - \Delta T_2 = T_2 - \lambda \Delta T_1 \end{cases} \tag{53}$$

Since the two row vectors in $T(s)$ are mutually linearly independent, so

$$\begin{cases} T_2 \neq \lambda T_1 \\ \hat{T}_2 \neq \lambda \hat{T}_1 \end{cases} \tag{54}$$

The above equations mean that $\hat{T}_1$ and $\hat{T}_2$ cannot be zero at the same time.

1) One of $\hat{T}_1$ and $\hat{T}_2$ is zero, it is no problem to assume that $\hat{T}_1$ is zero (that is, $\hat{T}_1 = 0, \hat{T}_2 \neq 0$). Perform the following row transformations as follows.

$$\begin{cases} T_1 = \Delta T_1 \\ \hat{T}_2 = T_2 - \Delta T_2 = T_2 - \lambda \Delta T_1 = T_2 - \lambda T_1 \end{cases} \tag{55}$$

Equations (55) mean that in the matrix $T(s)$, all rows except the second row are kept unchanged and a row transformation (denoted as $I(2,1;-\lambda)$, that is performing a row transformation of the first row multiplied by a factor of $(-\lambda)$ to superimpose on the second row) is only performed on the second row. The new matrix $T(s)$ after the row transformation is also invertible.

2) $\hat{T}_1 \neq 0, \hat{T}_2 \neq 0$

$$\begin{cases} \hat{T}_1 = (1-\gamma_1)T_1 - \gamma_2 T_2 \\ \hat{T}_2 = (1-\lambda\gamma_2)T_2 - \lambda\gamma_1 T_1 \end{cases} \tag{56}$$

(a) $1-\gamma_1 \neq 0$ or $1-\lambda\gamma_2 \neq 0$, it is no problem to assume that $1-\gamma_1 \neq 0$.

Equations (56) mean that in the matrix $T(s)$, all rows except the first row are kept unchanged and a series of row transformations in turn are performed on first row as follows.

$$\begin{cases} I(1; 1-\gamma_1) \\ I(1,2;-\gamma_2) \end{cases} \tag{57}$$

where $I(1; 1-\gamma_1)$ is performing a row transformation of the first row multiplied by a factor of $(1-\gamma_1)$.

Similarly, the new matrix $T(s)$ after the above row transformations is also invertible.

(b) $1-\gamma_1 = 0$ and $1-\lambda\gamma_2 = 0$, so $\gamma_1 \neq 0, \gamma_2 \neq 0$.

$$\hat{T}_1 = -\gamma_2 T_2 = -\frac{1}{\lambda}T_2 \tag{58}$$

Equation (58) means that in the matrix $T(s)$, all rows except the second row are kept unchanged and a row transformation is performed on the second row as follows.

$$I(2;-1/\lambda) \tag{59}$$

The second row of the new matrix $T(s)$ after the above row transformation is transformed into the row vector $\hat{T}_1$.

In turn, a series of row transformations are performed on the matrix $T(s)$ to eliminate the influence of the transfer functions of the mutually node-joint forward paths from two control inputs to two outputs though the same node and the invertibility of $T(s)$ is maintained until there are no transfer functions of the mutually node-joint forward paths from two control inputs to two outputs though the same node in $T(s)$.

It is then assumed that there are forward paths from the two different control inputs such as $u_1$ and $u_2$ to one output such as $y_i$ ($i = 1, 2$). Then, a series of the column transformations (that are similar to the above row transformations) are performed on the matrix $T(s)$ to eliminate the transfer functions of the forward paths from either $u_1$ or $u_2$ to output $y_i$ and the invertibility of $T(s)$ is maintained. In turn, a series of column transformations are performed on the matrix $T(s)$ to eliminate the transfer functions of the forward paths from the different control inputs to one output and the invertibility of $T(s)$ is maintained until there are no transfer functions of the forward paths from the different control inputs to the same output in $T(s)$.

Therefore, the matrix $T(s)$ is transformed into a new matrix whose each row (or column) has one and only one non-zero item. It means that there exists at least one general 2-path in the signal flow graph, and so Theorem 3.1 holds. □

The Theorem 3.1 is a sufficient condition on $F(z)$ or $R$ whether there exists a general 2-path in the signal flow graph.

**Theorem 3.2.** For a TITO standard signal flow graph and its system $\sum(A, B, C)$, if its transfer function $T(s)$ is invertible on $R$ (i.e., $rank(T(s)) = 2$ on $R$), then $T(s, z)$ is also invertible on $F(z)$ and there exists at least one general 2-path in the signal flow graph.

**Theorem 3.3.** For a TITO standard signal flow graph and its system $\sum(A, B, C)$, if $T(s) = 0$ on $R$ or $T(s, z) = 0$ on $F(z)$ (i.e., $rank(T(s, z)) = 0$), then there is no path between the two inputs and two outputs (called that there is general 0-path in the signal flow graph).

## 4. A necessary condition on $F(z)$

**Definition 4.1.** For a TITO standard signal flow graph and its system $\sum(A, B, C)$, if all paths from $u_i$ to $y_1$ or $y_2$ pass through a same node, then the node is called the reducible node of input $u_i$ (denoted as $v(u_i)$). If all paths from $u_1$ or $u_2$ to $y_i$ pass through a same node, then the node is called the reducible node of $y_i$ (denoted as $v(y_i)$). If all paths from $u_i$ to $y_j$ pass through one same node, then the node is called the reducible node of $(u_i, y_j)$ (denoted as $v(u_i, y_j)$).

**Definition 4.2.** For a TITO standard signal flow graph and its system $\sum(A, B, C)$, it is assumed that the reducible nodes of $(u_i, y_j)$ except $u_i$ and $y_j$ are $v_1(u_i, y_j)$, $v_2(u_i, y_j)$, $\cdots$, $v_{n_{ij}}(u_i, y_j)$ (referred as $v_1, v_2, \cdots, v_{n_{ij}}$) in sequence from $u_i$ to $y_j$. Then, $(u_i, y_j)$ can be decomposed into $(n_{ij} + 1)$ segments, each of whose input node and output node are its adjacent reducible nodes, so that the transfer function of $(u_i, y_j)$ can also be decomposed into the sub transfer functions (often referred as STFs) of $(n_{ij} + 1)$ segments and some sub transfer functions of global associative loop sets defined below. All segments of $(u_i, y_j)$ can be divided into two types: single-branch segment (called first-type segment, referred as SB) and non-single-branch segment which is composed of more than one sub-paths but does not include anyone SB sub-segment (called second-type segment, referred as NSB). It is obvious that each NSB segment is adjacent to two SB segments if exist.

**Definition 4.3.** For a TITO standard signal flow graph and its system $\sum(A, B, C)$, if an associative loop set is node-joint with one reducible node of $(u_i, y_j)$, it is called a first-type global associative loop set of $(u_i, y_j)$ (referred as FGALS of $(u_i, y_j)$). If at least one loop in an associative loop set is node-joint with all sub-paths in one NSB at the same time but is not node-joint with anyone reducible node of $(u_i, y_j)$, it is called a second-type global associative loop set of $(u_i, y_j)$ (referred as SGALS of $(u_i, y_j)$). The STF of each global associative loop set is called the impact factor of the global associative loop set (referred as the impact factor), which is an independent STF in the decomposition of the transfer function of $(u_i, y_j)$.

**Definition 4.4.** For a TITO standard signal flow graph and its system $\sum(A, B, C)$, if each loop in an associative loop set is not node-joint with all sub-paths in one NSB and is also not node-joint with anyone reducible node of $(u_i, y_j)$, it is called a local associative loop set of $(u_i, y_j)$ (referred as LALS of $(u_i, y_j)$). Then, the STF of each LALS (which is not an independent STF in the decomposition of the transfer function) is included in the STF of the corresponding NSB segment.

It is obvious that each FGALS of $(u_i, y_j)$ (such as $Q_k$) must branch-joint with (at least) one SB segment (such as first-order branch $x_\mu$) and it is easy to extract the impact factor of the FGALS by identifying the branch $x_\mu$. If the primary factor of $x_\mu$ is not an irreducible factor in the denominator polynomial and is also not included in the numerator polynomial of the transfer function of $(u_i, y_j)$ (denoted as $g(u_i, y_j)$), then the branch $x_\mu$ is a SB segment.

**Definition 4.5.** For a TITO standard signal flow graph and its system $\sum(A, B, C)$, if there is one FGALS of $(u_i, y_j)$ (such as $Q_k$) which branch-joint with one SB segment (such as first-order branch $x_\mu$), its sub-determinant is denoted as $\overline{\Delta}_k$ and the cofactor value of $\overline{\Delta}_k$ for the branch $x_\mu$ is denoted as $\overline{\Delta}_{k,}$, then the impact factor $g(Q_k)$ (with regard to its branch-joint SB segment) is given by

$$g(Q_k) = \overline{\Delta}_{k,} / \overline{\Delta}_k \triangleq N_k(s, z) / M_k(s, z) \quad (60)$$

After only the first-order branch $x_\mu$ is opened, the loop set $Q_k$ will be transformed into another loop set (called the downgraded loop set of $Q_k$, denoted as $\overline{Q}_{k,}$) whose sub-determinant is just $\overline{\Delta}_{k,}$. It is assumed that $\overline{Q}_{k,}$ is composed of $n_k$ node-disjoint associative loop sub-sets (called the downgraded associative loop sub-sets of $Q_k$, denoted as the set $\{Q_{k,\lambda}\}$) whose corresponding sub-determinants are denoted as the set $\{\overline{\Delta}_{k,\lambda}\}$. Then, the following equation holds.

$$\overline{\Delta}_{k,} = \begin{cases} \prod_{\lambda=1}^{n_k} \overline{\Delta}_{k,\lambda}, n_k \geq 1 \\ 1, n_k = 0 \end{cases} \quad (61)$$

The decomposition properties of the numerator polynomial $N_k(s, z)$ and the relationship between $N_k(s, z)$ and the set $\{Q_{k,\lambda}\}$ refer to Theorem 2.23.

As the same derivations as above, the impact factors of all FGALS of $(u_i, y_j)$ can be extracted from $g(u_i, y_j)$.

$$g(u_i, y_j) = g_1(u_i, y_j)(\prod_k g(Q_k)) \quad (62)$$

The above method to obtain the impact factor $g(Q_k)$ is also used to obtain the impact factor of one SGALS.

**Definition 4.6.** For a TITO standard signal flow graph and its system $\Sigma(A, B, C)$, if there is one SGALS of $(u_i, y_j)$ (such as $\bar{Q}_\sigma$) in which at least one loop is branch-joint with all sub-paths of one NSB segment (assumed it is the $\mu$th segment of $(u_i, y_j)$) at the same time, its sub-determinant is denoted as $\bar{\Delta}_\sigma$ and the cofactor value of $\bar{\Delta}_\sigma$ for all sub-paths of the $\mu$th segment (with the loops node-jointing all sub-paths of the $\mu$th segment at the same time removed) is denoted as $\bar{\Delta}_{\sigma,}$, then the impact factor $g(\bar{Q}_\sigma)$ (with regard to all sub-paths of the $\mu$th segment at the same time) is given by

$$g(\bar{Q}_\sigma) = \bar{\Delta}_{\sigma,} / \bar{\Delta}_\sigma \triangleq \bar{N}_\sigma(s,z) / \bar{M}_\sigma(s,z) \quad (63)$$

After only the loops in $\bar{Q}_\sigma$ which branch-joint with each sub-path of the $\mu$th segment at the same time are opened, the loop set $\bar{Q}_\sigma$ will be transformed into another loop set (called the downgraded loop set of $\bar{Q}_\sigma$, denoted as $\bar{Q}_{\sigma,}$) whose sub-determinant is just $\bar{\Delta}_{\sigma,}$. It is assumed that $\bar{Q}_{\sigma,}$ is composed of $\bar{n}_\sigma$ node-disjoint associative loop sub-sets (called the downgraded associative loop sub-sets of $\bar{Q}_\sigma$, denoted as the set $\{\bar{Q}_{\sigma,\upsilon}\}$) whose corresponding sub-determinants are denoted as the set $\{\bar{\Delta}_{\sigma,\upsilon}\}$. Then, the following equation holds.

$$\bar{\Delta}_{\sigma,} = \begin{cases} \prod_{\upsilon=1}^{\bar{n}_\sigma} \bar{\Delta}_{\sigma,\upsilon}, \bar{n}_\sigma \geq 1 \\ 1, \bar{n}_\sigma = 0 \end{cases} \quad (64)$$

The decomposition properties of the numerator polynomial $\bar{N}_\sigma(s,z)$ and the relationship between $\bar{N}_\sigma(s,z)$ and the set $\{\bar{Q}_{\sigma,\upsilon}\}$ also refer to Theorem 2.23.

As the same derivations as above, the impact factors of all SGALS of $(u_i, y_j)$ can be extracted from $g_1(u_i, y_j)$.

$$g_1(u_i, y_j) = g_2(u_i, y_j)(\prod_\sigma g(\bar{Q}_\sigma)) \quad (65)$$

Then, $g_2(u_i, y_j)$ is the product of the STFs of $(n_{ij}+1)$ segments. The STF of each SB segment is the gain of its first-order branch (whose numerator or denominator is one or the corresponding primary factor), which can uniquely point to the SB segment, and the numerator polynomial of the STF of each NSB segment is one irreducible polynomial about primary factors, which will be proved below. There is a one-to-one correspondence between each NSB segment which may contain some LALSs of $(u_i, y_j)$ and the numerator polynomial of its STF.

**Theorem 4.7.** For a TITO standard signal flow graph and its system $\Sigma(A, B, C)$, if a NSB segment of $(u_i, y_j)$ (whose input and output nodes are assumed to be $v_{k_1}$ and $v_{k_2}$) is $(v_{k_1}, v_{k_2})$ and the input branches of $v_{k_1}$ and the output branches of $v_{k_2}$ are all opened (such that there is not anyone FGALS in the NSB segment), the STF of $(v_{k_1}, v_{k_2})$ (which only the loops in possible SGALSs that branch-joint with each sub-path of the NSB segment at the same time are opened, such that there is no any SGALS in the NSB segment) is given by

$$g(v_{k_1}, v_{k_2}) = N(s,z) / M(s,z) \quad (66)$$

where $N(s,z)$ and $M(s,z)$ (i.e., the characteristic polynomial of the NSB segment $(v_{k_1}, v_{k_2})$) is coprime.

Then the numerator polynomial $N(s,z)$ is an irreducible polynomial about primary factors, such that $N(s,z)$ can uniquely point to the NSB segment (so that the input and output nodes of the NSB segment).

**Proof.** It is assumed that $N(s,z)$ is reducible about primary factors and $N(s,z) = N_1(s,z)N_2(s,z)$, where $N_1(s,z)$ and $N_2(s,z)$ are two polynomial about primary factors.

It is obvious that the irreducible polynomial factor (corresponding to anyone downgraded associative loop sub-set of a FGALS such as $Q_k$ or SGALS such as $\bar{Q}_\sigma$, which must have been transferred into the numerator polynomial of the impact factor $g(Q_k)$ or $g(\bar{Q}_\sigma)$ if existed before) and the primary factor of any first-order branch in the FGALS or SGALS (which must also have been transferred into the numerator polynomial of the impact factor $g(Q_k)$ or $g(\bar{Q}_\sigma)$ if existed before) are not multiplication factors of $N(s,z)$ (so that $N_1(s,z)$ or $N_2(s,z)$). Although the denominator polynomial $M(s,z)$ may include some irreducible polynomial factors (each of which corresponding to one downgraded associative loop sub-set of the FGALS or SGALS must also be the multiplication factor of the numerator polynomial of the impact factor $g(Q_k)$ or $g(\bar{Q}_\sigma)$), $N_2(s,z)$ is unchanged during the below opening operation on the branch $x(\bar{v})$ included in $N_1(s,z)$.

If $d^-(\bar{v}) = 1$ for one internal node $\bar{v}$ in the NSB segment, the unique input branch of the node $\bar{v}$ denoted as $x(\bar{v})$ (whose primary factor is denoted as $(s - z(\bar{v}))$ is assumed to be only included in $N_1(s,z)$ but not in $N_2(s,z)$ and a set of the branches (whose input nodes are the node $\bar{v}$) is denoted as $\{x_\kappa(\bar{v})\}$. After the opening operation on the branch $x(\bar{v})$ is performed, the primary factors of the branch $x(\bar{v})$ and all branches in $\{x_\kappa(\bar{v})\}$ will be not included in $N_3(s,z)$ and $M_3(s,z)$ as shown in (67), and there is at least one path in the remaining NSB segment.

$$\lim_{z(\bar{v}) \to \infty} \frac{N(s,z)}{M(s,z)} = (\lim_{z(\bar{v}) \to \infty} \frac{N_1(s,z)}{M(s,z)}) N_2(s,z) \triangleq \frac{N_3(s,z)}{M_3(s,z)} \neq 0$$
(67)

It means that the primary factors of the branch $x(\bar{v})$ and all branches in $\{x_\kappa(\bar{v})\}$ must be only included in $N_1(s,z)$ at the same time but not in $N_2(s,z)$. Otherwise, it is assumed that one branch $x_\kappa(\bar{v})$ is only included in $N_2(s,z)$ but not in $N_1(s,z)$, and $N_{2,\kappa}(s,z)$ (in which $(s - z_\kappa(\bar{v}))$ is only included) is the irreducible polynomial factor of $N_2(s,z)$.

If there is not anyone loop branch-joint with the branch $x_\kappa(\bar{v})$, the primary factor $(s - z_\kappa(\bar{v}))$ must be a polynomial factor of the denominator polynomial $M(s,z)$. When the opening operation on the branch $x(\bar{v})$ is performed, the primary factor $(s - z_\kappa(\bar{v}))$ included in the denominator $M(s,z)$ must be only eliminated by $N_{2,\kappa}(s,z)$. Therefore, there exist the expression: $N_{2,\kappa}(s,z) = (s - z_\kappa(\bar{v}))$, such that the primary factor $(s - z_\kappa(\bar{v}))$ must be one polynomial factor of $N(s,z)$ and $M(s,z)$ at the same time. It is contradictory to that $N(s,z)$ and $M(s,z)$ is coprime.

If there is at least one loop branch-joint with the branch $x_\kappa(\bar{v})$ (such that the branch $x(\bar{v})$) whose loop set is denoted as $Q_\kappa$ (in which each loop branch-joints with the branch $x_\kappa(\bar{v})$), the primary factors of the branch $x_\kappa(\bar{v})$ and the branch $x(\bar{v})$ must be only included in a same irreducible polynomial factor (denoted as $M_2(s,z)$, which corresponds to one unique associative loop set (denoted as $Q(x(\bar{v}))$) branch-joint with the branch $x(\bar{v})$) of the denominator $M(s,z)$. Through only the loops in $Q(x(\bar{v}))$ which branch-disjoint with the branch $x(\bar{v})$ are opened (it should be noted here that, this operation does not affect anyone path or other associative loop sets in the NSB segment), $Q(x(\bar{v}))$ will be transformed into another associative loop set denoted as $\tilde{Q}(x(\bar{v}))$ (whose one certain sub-set is just the set $Q_\kappa$) and $\tilde{Q}(x(\bar{v}))$ is one downgraded associative loop sub-set of $Q(x(\bar{v}))$ (in which each loop branch-joints with the branch $x(\bar{v})$), such that $M_2(s,z)$ is also transformed into another irreducible polynomial factor denoted as $\tilde{M}_2(s,z)$ and $N_1(s,z)$ (or $N_2(s,z)$) will be transformed into another polynomial factor $\tilde{N}_1(s,z)$ (or $\tilde{N}_2(s,z)$). $\tilde{N}_{2,\kappa}(s,z)$ (in which the primary factor $(s-z_\kappa(\bar{v}))$ is only included) is the irreducible factor of $\tilde{N}_2(s,z)$ and the primary factor $(s-z_\kappa(\bar{v}))$ will be only included in $\tilde{N}_2(s,z)$ and $\tilde{N}_{2,\kappa}(s,z)$ but not in $\tilde{N}_1(s,z)$.

The highest-order term about the primary factors in $\tilde{M}_2(s,z)$ (which only includes the product factor $(s-z(\bar{v}))$ in all additive terms of $\tilde{M}_2(s,z)$) is just the product of the primary factors of all first-order branch included in $\tilde{Q}(x(\bar{v}))$, and the highest-order term is only active in $\tilde{M}_2(s,z)$ when the opening operation on the branch $x(\bar{v})$ is performed. When the opening operation on the branch $x(\bar{v})$ is performed, the primary factor $(s-z_\kappa(\bar{v}))$ included in the denominator $\tilde{M}_2(s,z)$ must be only eliminated by $\tilde{N}_{2,\kappa}(s,z)$. Therefore, there exist the expression: $\tilde{N}_{2,\kappa}(s,z) = (s-z_\kappa(\bar{v}))$. It is further obtained that each loop in the set $Q_\kappa$ must be branch-joint with each path in $(v_{k_1}, v_{k_2})$, such that there exist at least one SGALS in $(v_{k_1}, v_{k_2})$. It is a contradiction.

Therefore, for each internal node whose in-degree is one in the NSB segment, the primary factors of its input branch and output branches must be all included either in $N_1(s,z)$ or in $N_2(s,z)$.

As the same derivation as above, for each internal node whose out-degree is one in the NSB segment, the primary factors of its input branches and output branch must be all included either in $N_1(s,z)$ or in $N_2(s,z)$. So, the NSB segment will be divided into two branch-disjoint parallel sub-segments (denoted as $(v_{k_1}, v_{k_2})_1$ and $(v_{k_1}, v_{k_2})_2$) whose STFs are denoted as $g(v_{k_1}, v_{k_2})_1$ and $g(v_{k_1}, v_{k_2})_2$, the set of all first-order branches included in $(v_{k_1}, v_{k_2})_1$ is denoted as $Z_1$ and the set of all first-order branches included in $(v_{k_1}, v_{k_2})_2$ is denoted as $Z_2$. The following equations hold.

$$g(v_{k_1}, v_{k_2}) = g(v_{k_1}, v_{k_2})_1 + g(v_{k_1}, v_{k_2})_2 \quad (68)$$

$$\begin{cases} z = Z_1 \cup Z_2 \\ Z_1 \cap Z_2 = \varnothing \end{cases} \quad (69)$$

$$g(v_{k_1}, v_{k_2})_1 \triangleq \bar{N}_1(s, Z_1)/M_1(s, Z_1) \quad (70)$$

$$g(v_{k_1}, v_{k_2})_2 \triangleq \bar{N}_2(s, Z_2)/M_2(s, Z_2) \quad (71)$$

$$g(v_{k_1}, v_{k_2}) = \frac{N_1(s, Z_1)N_2(s, Z_2)}{M_1(s, Z_1)M_2(s, Z_2)} \quad (72)$$

$$\frac{N_1(s,Z_1)N_2(s,Z_2)}{M_1(s,Z_1)M_2(s,Z_2)} = \frac{\bar{N}_1(s,Z_1)}{M_1(s,Z_1)} + \frac{\bar{N}_2(s,Z_2)}{M_2(s,Z_2)} \quad (73)$$

where $\bar{N}_1(s,Z_1)$ and $M_1(s,Z_1)$ is coprime, $\bar{N}_2(s,Z_2)$ and $M_2(s,Z_2)$ is also coprime.

$$N_1(s,Z_1) = \frac{M_2(s,Z_2)}{N_2(s,Z_2)}\bar{N}_1(s,Z_1) + \frac{\bar{N}_2(s,Z_2)}{N_2(s,Z_2)}M_1(s,Z_1) \quad (74)$$

$$\begin{cases} N_1(s,Z_1) \neq 0 \\ \bar{N}_1(s,Z_1) \neq 0 \\ M_1(s,Z_1) \neq 0 \end{cases} \quad (75)$$

Equation (74) is true for the indeterminate vectors $Z_1$ and $Z_2$ such that the following equations must hold.

$$\begin{cases} \dfrac{M_2(s,Z_2)}{N_2(s,Z_2)} = k_{21} \neq 0, k_{21} \in R \\ \dfrac{\bar{N}_2(s,Z_2)}{N_2(s,Z_2)} = k_{22} \neq 0, k_{22} \in R \end{cases} \quad (76)$$

$$\frac{\bar{N}_2(s,Z_2)}{M_2(s,Z_2)} = \frac{k_{22}}{k_{21}} \triangleq k_2 \neq 0, k_2 \in R \quad (77)$$

It is contradictory to that $\bar{N}_2(s,Z_2)$ and $M_2(s,Z_2)$ is coprime. So Theorem 4.7 holds. □

**Theorem 4.8.** For a TITO standard signal flow graph and its system $\Sigma(A,B,C)$, it is assumed that the reducible nodes of $(u_i, y_j)$ except $u_i$ and $y_j$ are $v_1, v_2, \cdots, v_{n_{ij}}$ in sequence from $u_i$ to $y_j$. Then the transfer function of $(u_i, y_j)$ can be uniquely decomposed into the STFs of $(n_{ij}+1)$ segments and the impact factors of all global associative loop sets.

$$\begin{cases} g(u_i, y_j) = g_2(u_i, y_j)(\prod_k g(Q_k))(\prod_\sigma g(\bar{Q}_\sigma)) \\ g_2(u_i, y_j) = \prod_{k=0}^{n_{ij}} g(v_k, v_{k+1}), v_0 \triangleq u_i, v_{n_{ij}+1} \triangleq y_j \end{cases} \quad (78)$$

**Theorem 4.9.** For a TITO standard signal flow graph and its system $\Sigma(A,B,C)$, its transfer function is given by

$$T(s,z) = \begin{bmatrix} T_{11} & T_{12} \\ T_{21} & T_{22} \end{bmatrix} \quad (79)$$

If $T(s,z) \neq 0$ and $\det(T(s,z)) \equiv 0$ (i.e., $rank(T(s,z))=1$), there is not general 2-path in the signal flow graph. Moreover, all paths from the two inputs to the two outputs intersect at the same node (called that there is a general 1-path in the TITO signal flow graph).

**Proof.** 1) If $T_{11} = T_{12} = 0$, all paths from the two inputs to the two outputs intersect at the output node $y_2$.

2) If $T_{21} = T_{22} = 0$, all paths from the two inputs to the two outputs intersect at the output node $y_1$.

3) If $T_{11} = T_{21} = 0$, all paths from the two inputs to the two outputs intersect at the input node $u_2$.

4) If $T_{12} = T_{22} = 0$, all paths from the two inputs to the two outputs intersect at the input node $u_1$.

5) For a general case that each element of $T(s,z)$ is not equal to zero, $rank(T(s,z))=1$ means that only one row is independent. It is assumed that the second row of $T(s,z)$ is independent.

$$\begin{cases} T_{11} = \beta T_{21} \\ T_{12} = \beta T_{22} \end{cases} \quad (80)$$

It is assumed that the reducible node of the input node $u_1$ which is furthest away from $u_1$ is $v$ and the STF of $(u_1, v)$ which includes the same impact factors of the common global associative loop sets (that are included in $T_{11}$ and $T_{21}$) is $g_1$.

$$\begin{cases} T_{11} \triangleq g_1 \overline{T}_{11} \\ T_{21} \triangleq g_1 \overline{T}_{21} \end{cases} \quad (81)$$

$$\beta = \overline{T}_{11} / \overline{T}_{21} \quad (82)$$

There is not any same segment between $(v, y_1)$ and $(v, y_2)$ (so that there is not any same STF for segment between $\overline{T}_{11}$ and $\overline{T}_{21}$). Because the denominator polynomial of the STF of each SB segment of $(v, y_1)$ or $(v, y_2)$ is the primary factor of its first-order branch which can uniquely point to the SB segment and the numerator polynomial of the STF of each NSB segment of $(v, y_1)$ or $(v, y_2)$ is an irreducible polynomial about primary factors which will also uniquely point to the NSB segment, the function $\beta$ can uniquely point to $\overline{T}_{11}$ and $\overline{T}_{21}$ such that $(v, y_1)$ and $(v, y_2)$. Therefore, the node $v$ must also be the reducible node of the input node $u_2$ which is furthest away from $u_2$ according to the second equation in (80). The STF of $(u_2, v)$ which includes the same impact factors of the common global associative loop sets (that are included in $T_{12}$ and $T_{22}$) is assumed to be $g_2$. The following equations hold.

$$\begin{cases} T_{12} \triangleq g_2 \overline{T}_{11} \\ T_{22} \triangleq g_2 \overline{T}_{21} \end{cases} \quad (83)$$

Therefore, all paths from the two inputs to the two outputs intersect at the same node $v$, which is both the reducible node of $u_1$ which is furthest away from $u_1$ and the reducible node of $u_2$ which is furthest away from $u_2$. That is to say, there is not general 2-path in the signal flow graph. □

It is obvious that for the above node $v$ with respect to the two outputs, there are three possible cases: 1) $v = y_1$, $\Rightarrow$ $\overline{T}_{11} = 1$; 2) $v = y_2$, $\Rightarrow$ $\overline{T}_{21} = 1$; 3) a general case, $v \neq y_1$ and $v \neq y_2$.

For the above node $v$ with respect to the two inputs, there are also three possible cases: 1) $v = u_1$, $\Rightarrow$ $g_1 = 1$; 2) $v = u_2$, $\Rightarrow$ $g_2 = 1$; 3) a general case, $v \neq u_1$ and $v \neq u_2$.

**Theorem 4.10.** For a TITO standard signal flow graph and its system $\sum(A, B, C)$, if there is at least one general 2-path in the signal flow graph, then $rank(T(s, z)) = 2$ on $F(z)$ (which is a necessary condition on $F(z)$ whether there exists a general 2-path in the signal flow graph).

## 5. A sufficient and necessary condition on $F(z)$

Finally, according to the above analysis, it is easy to obtain a sufficient and necessary condition on $F(z)$ whether there exists a general 2-path in a TITO signal flow graph.

**Theorem 5.1.** For a TITO standard signal flow graph and its system $\sum(A, B, C)$, the following conclusions hold.

1) There is at least one general 2-path in the signal flow graph if and only if $rank(T(s, z)) = 2$ on $F(z)$;

2) There is a general 1-path in the signal flow graph if and only if $rank(T(s, z)) = 1$ on $F(z)$;

3) There is general 0-path in the signal flow graph if and only if $rank(T(s, z)) = 0$ on $F(z)$ (i.e., $T(s, z) \equiv 0$).

Therefore, there are only three cases (i.e., either the case where there is at least one general 2-path, or the case where there is a general 1-path, or the case where there is general 0-path) in a TITO signal flow graph. And it must be one of the three cases.

## 6. An equivalent necessary condition on $R$

For one value $\alpha$ ($\alpha \triangleq [\alpha_1, \alpha_2, \cdots, \alpha_n]^T$) that $rankT(s, \alpha) = 1$, then either $\alpha \in \Phi_0$ (as shown in (4), which the function $f(z)$ is replaced by $\det(T(s, \alpha))$) or $\alpha \in \Phi_1$ (as shown in (5), which the function $f(z)$ is replaced by $\det(T(s, \alpha))$), that is to say, either there is a general 1-path or there is at least one general 2-path in a TITO SFG.

It is assumed that $z \in \Phi_1$, that is, $rank(T(s, z)) = 1$ but there is at least one general 2-path in a TITO SFG (denoted as $D$), and it is obvious that each element of $T(s, z)$ is nonzero. Then, $\det(T(s, z)) = 0$ will be expressed as the expansion form of series and the following equation first holds from equation(79).

$$T_{11} T_{22} = T_{12} T_{21} \quad (84)$$

It is assumed that one general 2-path (whose total length is the shortest, denoted as $l_m$) defined as shortest general 2-path is two general paths such as $P_1$ and $P_2$ (whose lengths are $l_{m1}$ and $l_{m2}$ respectively). For any two paths such as $path(u_1, y_1)$ and $path(u_2, y_2)$ (which are branch-joint with the branch $x_k$) whose total length (denoted as $l$, $l \leq l_m$) is not greater than $l_m$, then a TITO sub SFG of $D$ (denoted as $D_k$) is composed of the above two paths (such that it is impossible for the case where there is general 0-path) and the total number (denoted as $l'$, $l' \leq l - 1 \leq l_m - 1$) of all branches in $D_k$ must be less than $l_m$, such that there is no general 2-path (otherwise, the total length of one general 2-path in $D_k$ will be less than $l_m$, and it is a contradiction) and all paths between the two inputs and two outputs in $D_k$ must be node-joint at one same node (denoted as $v$). All paths through the node $v$ in $D_k$ are divided into paths $(u_i, v, y_j)$, $i, j = 1, 2$, and their STFs in $D$ are denoted as $g(u_i, v, y_j)$, $i, j = 1, 2$. Therefore, the following equation holds.

$$g(u_1, v, y_1) g(u_2, v, y_2) = g(u_1, v, y_2) g(u_2, v, y_1) \quad (85)$$

Subtract (85) from (84) to obtain a new equation (still denoted as equation (84)) until the all cases that the total length of each two branch-joint paths such as $path(u_1, y_1)$ and $path(u_2, y_2)$ (or $path(u_1, y_2)$ and $path(u_2, y_1)$) is not greater than $l_m$ are eliminated from equation (84), such that there are only general 2-path cases (whose total lengths are $l_m$) and remaining two-path cases (whose total lengths are greater than $l_m$) in the updated equation (84).

For one general 2-path (whose total length is $l_m$) with two general paths such as $P_1$ and $P_2$ (assumed that they are included in $(u_1, y_1)$ and $(u_2, y_2)$ respectively), the transfer functions of $P_1$ and $P_2$ are denoted as $g(P_1)$ and $g(P_2)$ respectively.

$$g(P_1) = \frac{f_0^1(z)}{s^{l_{m1}}} + \frac{f_{11}^1(z) + f_{10}^1(z)}{s^{l_{m1}+1}} + \frac{f_{22}^1(z) + f_{21}^1(z) + f_{20}^1(z)}{s^{l_{m1}+2}} + \cdots \quad (86)$$

$$g(P_2) = \frac{f_0^2(z)}{s^{l_{m2}}} + \frac{f_{11}^2(z) + f_{10}^2(z)}{s^{l_{m2}+1}} + \frac{f_{22}^2(z) + f_{21}^2(z) + f_{20}^2(z)}{s^{l_{m2}+2}} + \cdots \quad (87)$$

$$g(P_1)g(P_2) = \frac{f_0(z)}{s^{l_m}} + \frac{f_{11}(z) + f_{10}(z)}{s^{l_m+1}} + \frac{f_{22}(z) + f_{21}(z) + f_{20}(z)}{s^{l_m+2}} + \cdots \quad (88)$$

$$\begin{cases} f_0^1(z) = 1 \\ f_0^2(z) = 1 \end{cases} \quad (89)$$

$$\begin{cases} f_0(z) = f_0^1(z) f_0^2(z) = 1 \\ f_{11}(z) = f_{11}^2(z) + f_{11}^1(z) \\ f_{22}(z) = f_{11}^1(z) f_{11}^2(z) + f_{22}^2(z) + f_{22}^1(z) \end{cases} \quad (90)$$

It is obvious that $f_0(z)$ is the number of all shortest general 2-path case, $f_{11}(z)$ is the ESPP of all first-order branches in the shortest general 2-path case, and $f_{22}(z)$ is the ESQP of all first-order branches in the shortest general 2-path case. Therefore, the series expansion form of the right and left side of the updated equation (84) will be obtained by use of linear superposition principle as follows.

$$\begin{cases} T_{11}T_{22} = \frac{\bar{f}_0(z)}{s^{l_m}} + \frac{\bar{f}_{11}(z) + \bar{f}_{10}(z)}{s^{l_m+1}} + \frac{\bar{f}_{22}(z) + \bar{f}_{21}(z) + \bar{f}_{20}(z)}{s^{l_m+2}} + \cdots \\ T_{12}T_{21} = \frac{\hat{f}_0(z)}{s^{l_m}} + \frac{\hat{f}_{11}(z) + \hat{f}_{10}(z)}{s^{l_m+1}} + \frac{\hat{f}_{22}(z) + \hat{f}_{21}(z) + \hat{f}_{20}(z)}{s^{l_m+2}} + \cdots \end{cases} \quad (91)$$

$$\begin{cases} \bar{f}_2(z) = \bar{f}_{22}(z) + \bar{f}_{21}(z) + \bar{f}_{20}(z) \\ \hat{f}_2(z) = \hat{f}_{22}(z) + \hat{f}_{21}(z) + \hat{f}_{20}(z) \end{cases} \quad (92)$$

Then $\bar{f}_0(z)$ (or $\hat{f}_0(z)$) is the total number of all shortest general 2-path cases (each of which is uniquely corresponding to one shortest general 2-path included in $(u_1, y_1)$ and $(u_2, y_2)$ (or $(u_1, y_2)$ and $(u_2, y_1)$) respectively), $\bar{f}_{11}(z)$ (or $\hat{f}_{11}(z)$) is the sum of the ESPPs of the shortest general 2-path cases, and $\bar{f}_{22}(z)$ (or $\hat{f}_{22}(z)$) is the sum of the ESQPs of the shortest general 2-path cases. As the same derivation as above, the following theorem holds.

**Theorem 6.1.** For a TITO standard signal flow graph and $z \in \Phi_1$, there are $n_s$ shortest general 2-path cases (whose order is $l_m$) if and only if the function $\bar{f}_{22}(z)$ (or $\hat{f}_{22}(z)$) (as shown in (91)) on $F(z)$ can be uniquely decomposed into $n_s$ ESQPs (as shown in (43)), each of which uniquely corresponds to one unique shortest general 2-path included in $(u_1, y_1)$ and $(u_2, y_2)$ (or $(u_1, y_2)$ and $(u_2, y_1)$). Therefore, the function $\bar{f}_{22}(z)$ (or $\hat{f}_{22}(z)$) on $F(z)$ uniquely points to each shortest general 2-path such that also uniquely points to the input nodes and output nodes.

In order to easily derive the conclusions needed, the SFG of $D$ is modified to another SFG (denoted as $\bar{D}$) as shown in Fig.3. In the modified SFG, all paths through the nodes $y_1$ (or $y_2$) and $u_2$ in turn from $u_1$ to $y_2$ (or $y_1$) are denoted as $(u_1, y_1(u_2), y_2)$ (or $(u_1, y_2(u_2), y_1)$) and their transfer function is $g(1,1(2),2)$ (or $g(1,2(2),1)$). There exists one path in $(u_1, y_1(u_2), y_2)$ (or $(u_1, y_2(u_2), y_1)$) if and only if there is one corresponding general 2-path. Therefore, the following equations hold.

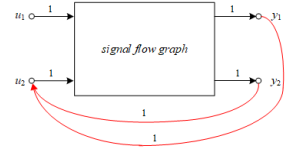

Fig.3. The modified signal flow graph

$$\begin{cases} g(1,1(2),2) = \frac{\bar{g}_0(z)}{s^{l_m}} + \frac{\bar{g}_{11}(z) + g_{10}(z)}{s^{l_m+1}} + \frac{\bar{g}_{22}(z) + \bar{g}_{21}(z) + \bar{g}_{20}(z)}{s^{l_m+2}} + \cdots \\ g(1,2(2),1) = \frac{\hat{g}_0(z)}{s^{l_m}} + \frac{\hat{g}_{11}(z) + \hat{g}_{10}(z)}{s^{l_m+1}} + \frac{\hat{g}_{22}(z) + \hat{g}_{21}(z) + \hat{g}_{20}(z)}{s^{l_m+2}} + \cdots \end{cases} \quad (93)$$

$$\begin{cases} \bar{f}_0(z) = \bar{g}_0(z) \\ \hat{f}_0(z) = \hat{g}_0(z) \end{cases} \quad (94)$$

$$\begin{cases} \bar{f}_{11}(z) = \bar{g}_{11}(z) \\ \hat{f}_{11}(z) = \hat{g}_{11}(z) \end{cases} \quad (95)$$

$$\begin{cases} \bar{f}_{22}(z) = \bar{g}_{22}(z) \\ \hat{f}_{22}(z) = \hat{g}_{22}(z) \end{cases} \quad (96)$$

So, the following derivations on $F(z)$ hold.

$$\begin{cases} rank(T(s,z)) = 1 \Leftrightarrow T_{11}T_{22} = T_{12}T_{21} \\ T_{11}T_{22} = T_{12}T_{21} \Leftrightarrow \bar{f}_2(z) = \hat{f}_2(z) \\ T_{11}T_{22} = T_{12}T_{21} \Leftrightarrow \bar{f}_{22}(z) = \hat{f}_{22}(z) \\ \bar{f}_{22}(z) = \hat{f}_{22}(z) \Leftrightarrow \bar{g}_{22}(z) = \hat{g}_{22}(z) \\ \bar{g}_{22}(z) = \hat{g}_{22}(z) \Leftrightarrow g(1,1(2),2) = g(1,2(2),1) \\ g(1,1(2),2) = g(1,2(2),1) \Leftrightarrow y_1 = y_2 \end{cases} \quad (97)$$

A set of points with values of $z$ (denoted as $\Phi_3$) will be selected to obtain an equivalent necessary condition on $R$.

1) One sub-set (denoted as $\Phi_{31}$) of the set $\Phi_3$, which has $(2n+1)$ values of $z$.

(1) $z_i = ni$;
(2) $z_i = ni, i \neq k; z_k = nk - 1, k = 1, 2, \cdots, n$, i.e., it is a small perturbation process for only one column of value $z$ at one time and the total perturbation number is $n$ (that is, $n$ values of $z$);
(3) $z_i = ni, i \neq k; z_k = nk + 1, k = 1, 2, \cdots, n$.

2) $(n-1)$ sub-sets (denoted as $\Phi_{3j}$, $2 \leq j \leq n$), each of which has $(n+1)$ values of $z$.

(4) $z_i = (ni)^j$;
(5) $z_i = (ni)^j, i \neq k; z_k = (nk)^j - 1, k = 1, 2, \cdots, n$.

$$\Phi_3 = \bigcup_{j=1}^{n} \Phi_{3j} \quad (98)$$

Then there are $(n^2 + 2n)$ elements in the set $\Phi_3$.

If $rank(T(s,z)) = 1$ for $\forall z \in \Phi_3$, then either $\Phi_3 \subseteq \Phi_1$ or $\Phi_3 \subseteq \Phi_0$. It is assumed that $\Phi_3 \subseteq \Phi_1$.

Let

$$\begin{cases} \bar{f}_2(z) \triangleq \bar{\beta}_0 + \sum_{1 \leq i \leq n} \bar{\beta}_{i0} z_i + \sum_{1 \leq i \leq j \leq n} \bar{\beta}_{ij} z_i z_j \\ \hat{f}_2(z) \triangleq \hat{\beta}_0 + \sum_{1 \leq i \leq n} \hat{\beta}_{i0} z_i + \sum_{1 \leq i \leq j \leq n} \hat{\beta}_{ij} z_i z_j \end{cases} \quad (99)$$

Let $z \in \Phi_{31}$,

$$\frac{\partial^2}{\partial z_i^2}\bar{f}_2(z) = \frac{\partial^2}{\partial z_i^2}\hat{f}_2(z), z_i = ni \quad (100)$$

$$\bar{\beta}_{ii} = \hat{\beta}_{ii}, i = 1, 2, \cdots, n \quad (101)$$

Let $z \in \Phi_3$,

$$\frac{\partial}{\partial z_i}\bar{f}_2(z) = \frac{\partial}{\partial z_i}\hat{f}_2(z), z_i = (ni)^j, j = 1, 2, \cdots, n \quad (102)$$

$$\bar{\beta}_{i0} + \sum_{j \neq i}\bar{\beta}_{ij}z_j = \hat{\beta}_{i0} + \sum_{j \neq i}\hat{\beta}_{ij}z_j, z_i = (ni)^j \quad (103)$$

$$\bar{\beta}_{ij} = \bar{\beta}_{ji}, i, j = 1, 2, \cdots, n \quad (104)$$

$$\hat{\beta}_{ij} = \hat{\beta}_{ji}, i, j = 1, 2, \cdots, n \quad (105)$$

$$\begin{cases} \bar{\beta}_i \triangleq \left[\bar{\beta}_{i0}, \bar{\beta}_{i1}, \cdots, \bar{\beta}_{i(i-1)}, \bar{\beta}_{i(i+1)}, \cdots, \bar{\beta}_{in}\right]^T \\ \hat{\beta}_i \triangleq \left[\hat{\beta}_{i0}, \hat{\beta}_{i1}, \cdots, \hat{\beta}_{i(i-1)}, \hat{\beta}_{i(i+1)}, \cdots, \hat{\beta}_{in}\right]^T \end{cases} \quad (106)$$

$$D_0 \triangleq diag\left[1, n, \cdots, n(i-1), n(i+1), \cdots, n^2\right] \quad (107)$$

$$V \triangleq \begin{bmatrix} 1 & 1 & \cdots & 1 & 1 & \cdots & 1 \\ 1 & n & \cdots & n(i-1) & n(i+1) & \cdots & n^2 \\ \vdots & \vdots & \ddots & \vdots & \vdots & \ddots & \vdots \\ 1 & n^{n-1} & \cdots & (n(i-1))^{n-1} & (n(i+1))^{n-1} & \cdots & n^{2(n-1)} \end{bmatrix} \quad (108)$$

$$VD_0\bar{\beta}_i = VD_0\hat{\beta}_i \quad (109)$$

The matrix $V$ is one $n \times n$ Vandermonde matrix.

$$\bar{\beta}_i = \hat{\beta}_i, i = 1, 2, \cdots, n \quad (110)$$

$$\begin{cases} \bar{\beta}_{i0} = \hat{\beta}_{i0}, i = 1, 2, \cdots, n \\ \bar{\beta}_{ij} = \hat{\beta}_{ij}, i, j = 1, 2, \cdots, n \end{cases} \quad (111)$$

$$\bar{f}_2(z) = \hat{f}_2(z), z_i = ni \Rightarrow \bar{\beta}_0 = \hat{\beta}_0 \quad (112)$$

Therefore, if $rank(T(s, z)) = 1$ for $\forall z \in \Phi_3$ and $\Phi_3 \subseteq \Phi_1$, the following equation on $F(z)$ holds (that is, the derivations (97) on $F(z)$ also hold).

$$\bar{f}_2(z) = \hat{f}_2(z) \quad (113)$$

Combine (113) and (97), and obtain

$$\begin{cases} \bar{f}_{22}(z) = \hat{f}_{22}(z) \\ \bar{g}_{22}(z) = \hat{g}_{22}(z) \end{cases} \quad (114)$$

$$y_1 = y_2 \quad (115)$$

It means that both the output nodes $y_1$ and $y_2$ are the same node. It is a contradiction.

Therefore, $\Phi_3 \not\subset \Phi_1$ and $\Phi_3 \subset \Phi_0$. If $rank(T(s, \alpha)) = 1$ for $\forall \alpha \in \Phi_3$, then $rank(T(s, z)) = 1$ on $F(z)$, such that $rank(T(s, \alpha)) = 1$ ($\forall \alpha \in \Phi_3$) if and only if $rank(T(s, z)) = 1$ on $F(z)$ combined the case where if $rank(T(s, \alpha)) = 2$ ($\exists \alpha \in \Phi_3$), then $rank(T(s, z)) = 2$ on $F(z)$ (see Theorem 2.27) (or if $rank(T(s, z)) = 1$ on $F(z)$, that is, $det(T(s, z)) \equiv 0$, then $rank(T(s, \alpha)) = 1$ for $\forall \alpha \in \Phi_3$).

**Definition 6.2.** For a TITO standard signal flow graph and its system $\sum(A, B, C)$, the rank of the system is given by

$$r \triangleq \max\left\{rank(T(s, \alpha))|\forall \alpha \in \Phi_3\right\} \quad (116)$$

**Theorem 6.3.** For a TITO standard signal flow graph and its system $\sum(A, B, C)$, if there is at least one general 2-path in the signal flow graph, then $r = 2$ (which is an equivalent necessary condition on $R$ whether there exists a general 2-path in the signal flow graph).

## 7. An equivalent sufficient and necessary condition

Finally, combining the sufficient condition (see Section 3) and equivalent necessary condition (as shown in Section 6) on $R$, it is easy to obtain an equivalent sufficient and necessary condition on $R$ whether there exists a general 2-path in a TITO signal flow graph.

**Theorem 7.1.** For a TITO standard signal flow graph and its system $\sum(A, B, C)$, the following conclusions hold.

1) There is at least one general 2-path in the signal flow graph if and only if $r = 2$ on $R$;

2) There is a general 1-path in the signal flow graph if and only if $r = 1$ on $R$;

3) There is general 0-path in the signal flow graph if and only if $r = 0$ on $R$.

## 8. Computational complexity of the algorithm for the equivalent sufficient and necessary condition on $R$

In order to estimate the computational complexity of the algorithm for the equivalent sufficient and necessary condition on $R$, this algorithm is first given as follows.

**Proposition 8.1.** For a TITO standard signal flow graph and its system $\sum(A, B, C)$ as shown in equations (8), then for $\alpha \in \Phi_3$, the following equations hold (W. A. Wolovich, 1974; F. S. Pang, 1992).

$$T(s, \alpha) = C(sI - A)^{-1}B \quad (117)$$

$$(sI - A)^{-1} = \frac{R_1 s^{n-1} + R_2 s^{n-2} + \cdots + R_n}{s^n + a_1 s^{n-1} + \cdots + a_n}, a_0 = 1, R_0 = 0 \quad (118)$$

$$\begin{cases} R_k = AR_{k-1} + a_{k-1}I, k = 1, 2, \cdots, n \\ a_k = -\frac{1}{k}tr(AR_k), k = 1, 2, \cdots, n \end{cases} \quad (119)$$

$$\begin{cases} R_0 = 0 \\ R_1 = I \\ R_2 = A + a_1 I \\ R_3 = A^2 + a_1 A + a_2 I \\ \vdots \\ R_n = A^{n-1} + a_1 A^{n-2} + \cdots + a_{n-1}I \end{cases} \quad (120)$$

$$\Delta(s) = s^n + a_1 s^{n-1} + \cdots + a_n \quad (121)$$

$$N(s, \alpha) = \Delta(s)T(s, \alpha) = \sum_{k=1}^{n}(CR_k B)s^{n-k} \triangleq \sum_{k=1}^{n}\bar{R}_k s^{n-k} \quad (122)$$

$$N(s, \alpha) \triangleq \begin{bmatrix} N_{11} & N_{12} \\ N_{21} & N_{22} \end{bmatrix} \quad (123)$$

If $N(s, \alpha) = 0$

$$\begin{cases} N(s, \alpha) = 0 \Leftrightarrow N(s, z) = 0 \\ N(s, z) = 0 \Leftrightarrow rank(T(s, z)) = 0 \\ rank(T(s, z)) = 0 \Leftrightarrow r = 0 \end{cases} \quad (124)$$

If $N(s, \alpha) \neq 0$

$$\delta(s, \alpha) \triangleq N_{11}N_{22} - N_{12}N_{21} \quad (125)$$

$$\begin{cases} \delta(s,\alpha) = 0 \Leftrightarrow rank(T(s,\alpha)) = 1 \\ \delta(s,\alpha) \neq 0 \Leftrightarrow rank(T(s,\alpha)) = 2 \end{cases} \quad (126)$$

$$r = \max\{rank(T(s,\alpha)) | \forall \alpha \in \Phi_3\} \quad (127)$$

In the matrix $A$, its diagonal elements are the open poles of the first-order branches in the SFG and the remaining elements are either one or zero, such that the maximum absolute value (denoted as $|a_{\max}|$) of the coefficients $\{a_i\}_n$ in its characteristic polynomial will first be estimated. Because each element in the matrix $B$ (or $C$) is either one or zero, then the maximum absolute element (denoted as $|R_{\max}|$) of the coefficient matrix $R_k$ in equation (118) is also estimated so that the coefficient matrices (that is, the maximum absolute element (denoted as $|\bar{R}_{\max}|$) of the coefficient matrix $\bar{R}_k$) for Laplace operator $s$ can be estimated in equation (122). In this way, the computational complexity of the above algorithm with regard to the size of the SFG (i.e., the total number of all first-order branches in the SFG, which is $n$) can be obtained.

The maximum absolute value of the elements in $A$ is $(n^2)^n$, so there are the following estimations.

$$|a_{\max}| \leq 2(\prod_{i=1}^{n}(ni)^n) < 2n^{2n^2} \quad (128)$$

$$|R_{\max}| \leq n|a_{\max}| < 2n^{2n^2+1} \quad (129)$$

$$|\bar{R}_{\max}| \leq n|R_{\max}| < 2n^{2n^2+2} \quad (130)$$

For the above algorithm, the computational burden of elementary operations is mainly concentrated in equations (119), (122) and (125). For one rank calculation of the transfer function matrix, the estimated number of 16-bit by 16-bit multiplication operations for calculating the matrix $R_k$ in (119) is denoted as $F_{1k}(m)$, the estimated number of addition operations of 16-bit plus 16-bit to calculate the matrix $R_k$ is denoted as $F_{1k}(a)$, and the total estimated number of operations to calculate the matrix $R_k$ is denoted as $F_{1k}$. As the same definitions as above, the estimated numbers of operations for calculating the coefficient $a_k$ in (119) are denoted as $F_{2k}(m)$, $F_{2k}(a)$ and $F_{2k}$ respectively; the estimated numbers of operations for calculating the coefficient $\bar{R}_k$ in (122) are denoted as $F_{3k}(a)$ and $F_{3k}$ respectively; and the estimated numbers of operations for calculating $\delta(s,\alpha)$ in (125) are denoted as $F_4(m)$, $F_4(a)$ and $F_4$ respectively.

$$F_{1k}(m) \leq 2n^2(\frac{1}{16}\log_2(n^{2n}) \times \frac{1}{16}\log_2|R_{\max}| \times n)$$
$$< \frac{n^3}{128}(2n\log_2 n \times ((2n^2+1)\log_2 n + 1)) < \frac{n^6}{16}\log_2^2 n \quad (131)$$

$$F_{1k}(a) \leq \frac{1}{2}F_{1k}(m)(F_{1k}(m)+1)$$
$$< \frac{n^6}{32}\log_2^2 n(\frac{n^6}{16}\log_2^2 n + 1) < \frac{n^{12}}{256}\log_2^4 n \quad (132)$$

$$F_{1k} = F_{1k}(m) + F_{1k}(a) < \frac{n^{12}}{128}\log_2^4 n \quad (133)$$

$$F_1 \triangleq \sum_{k=1}^{n} F_{1k} < \frac{n^{13}}{128}\log_2^4 n \quad (134)$$

$$\begin{cases} F_{2k}(m) < \frac{n^6}{16}\log_2^2 n \\ F_{2k}(a) < \frac{n^{12}}{256}\log_2^4 n \\ F_{2k} < \frac{n^{12}}{128}\log_2^4 n \end{cases} \quad (135)$$

$$F_2 \triangleq \sum_{k=1}^{n} F_{2k} < \frac{n^{13}}{128}\log_2^4 n \quad (136)$$

$$F_{3k}(a) \leq 2n(\frac{n}{16}\log_2|\bar{R}_{\max}| + \frac{n}{16}\log_2(n|\bar{R}_{\max}|))$$
$$< \frac{n^2}{8}(\log_2(2n^{2n^2+2}) + \log_2(2n^{2n^2+3})) < n^4\log_2 n \quad (137)$$

$$F_{3k} < n^4\log_2 n \quad (138)$$

$$F_3 \triangleq \sum_{k=1}^{n} F_{3k} < n^5\log_2 n \quad (139)$$

$$F_4(m) \leq 2n^2(\frac{1}{16}\log_2|\bar{R}_{\max}| \times \frac{1}{16}\log_2|\bar{R}_{\max}|)$$
$$< \frac{n^2}{128}((2n^2+2)\log_2 n + 1)^2 < \frac{n^6}{8}\log_2^2 n \quad (140)$$

$$F_4(a) \leq \frac{1}{2}F_4(m)(F_4(m)+1)$$
$$< \frac{n^6}{16}\log_2^2 n(\frac{n^6}{8}\log_2^2 n + 1) < \frac{n^{12}}{64}\log_2^4 n \quad (141)$$

$$F_4 < \frac{n^{12}}{32}\log_2^4 n \quad (142)$$

The total estimated number of operations for one rank calculation of the transfer function matrix is denoted as $F_0$.

$$F_0 = F_1 + F_2 + F_3 + F_4 < \frac{n^{13}}{32}\log_2^4 n \quad (143)$$

Then, the total estimated number of operations for $(n^2+2n)$ rank calculations of the transfer function matrix is denoted as $F$.

$$F \leq (n^2+2n)F_0 < \frac{n^{15}}{16}\log_2^4 n \quad (144)$$

Therefore, the computational complexity of the algorithm for the equivalent sufficient and necessary condition on $R$ is $O(n^{15}\log_2^4 n)$. It is means that the general 2-path problem (which is a NPC decision problem, denoted as $S$) is a P problem such that $S \in P$ and then $P = NP$.

## 9. Conclusions

This paper has first established the necessary and sufficient condition on $F(z)$ for a general 2-path problem which is a NPC decision problem. Moreover, an equivalent sufficient and necessary condition on $R$ whether there exists a general 2-path is further obtained. Finally, the computational complexity of the algorithm for this equivalent sufficient and necessary condition means that the general 2-path problem is a P problem.

In this paper, one non-constructive proof method (i.e., existence proof method) is used to prove that the general 2-path problem is a P problem. Therefore, one constructive proof for P versus NP is still an open problem.

**Acknowledgements**

The author would like to thank Professor Wen Dingdou and Dr. Zhang Yang for their helpful discussion while preparing this work, and would like to especially thank Dr. Zhang Yang for drawing all figures in this paper.